\documentclass[english,a4paper]{article}
\usepackage[a4paper,margin=1in]{geometry}

\usepackage[dvips]{epsfig}
\usepackage{graphicx}
\usepackage{subfig}
\usepackage{amsmath}
\usepackage{amsfonts}
\usepackage{amssymb}
\usepackage{siunitx}
\usepackage{helvet}
\usepackage{color}
\usepackage[nomain,acronym]{glossaries}
\usepackage{authblk}
\newcommand{\comment}[1]{}

\newcommand{\retL}{\alpha}
\newcommand{\perL}{\beta}
\newcommand{\bfreq}{\omega}

\newcommand{\totsimtime}{T}
\newcommand{\growthspeed}{v}

\newcommand{\Rg}{R_\text{g}}
\newcommand{\Vg}{V_\text{g}}
\newcommand{\dc}{d_\text{c}}
\newcommand{\db}{d_\text{b}}
\newcommand{\df}{d_\text{f}}
\newcommand{\lac}{\Lambda}
\newcommand{\lacscale}{B_\text{$\Lambda$}}
\newcommand{\hittingscale}{B_\text{H}}
\newcommand{\meanL}{\langle L \rangle}

\newcommand{\mm}{\,\mu\text{m}}
\newcommand{\minute}{\,\text{min}^{-1}}

\newcommand{\Term}{Definition of Morphometrics}

\begin{document}
%\title{Fly Class IV Sensory Neurons Morphologically Encode the Nociceptive Function by Controlling the Branching and Self-Avoidance.}
%\title{The Morphology of the Fly Larva Class IV Sensory Neurons Can be Accounted for by Constant Branching and Self-avoidance}
\title{Morphology of Fly Larval Class IV Dendrites Accords with a Random Branching and Contact Based Branch Deletion Model.}
\author[1]{Sujoy Ganguly}%
\author[1]{Olivier Trottier}
\author[2]{Xin Liang}%
\author[1]{Hugo Bowne-Anderson}%
\author[1]{Jonathon Howard}
 \affil[1]{Department of Molecular Biophysics and Biochemistry, Yale University, New Haven}%
 \affil[2]{Tsinghua-Peking Joint Center for Life Sciences, School of Life Sciences, Tsinghua University, Beijing, China}
\date{\today}
\maketitle
\begin{abstract}
Dendrites are branched neuronal processes that receive input signals from other neurons or the outside world \cite{Stuart:2007nx}. 
To maintain connectivity as the organism grows, dendrites must also continue to grow.
For example, the dendrites in the peripheral nervous system continue to grow and branch to maintain proper coverage of their receptor fields \cite{Bentley:1981kq,Truman:1988rw, Li:2005vn, Hitchcock:1987sf}.
One such neuron is the \emph{Drosophila melanogaster} class IV dendritic arborization neuron \cite{Parrish:2009pb}.
The dendritic arbors of these neurons tile the larval surface \cite{Bodmer1987}, where they detect localized noxious stimuli, such as jabs from parasitic wasps \cite{Hwang:2007qq}. % lying in a shallow space in the epidermis between the cuticle and the underlying basement membrane. 
In the present study, we used a novel measure, the hitting probability, to show that the class IV neuron forms a tight mesh that covers the larval surface.
Furthermore, we found that the mesh size remains largely unchanged during the larval stages, despite a dramatic increase in overall size of the neuron and the larva.
We also found that the class IV dendrites are dense (assayed with the fractal dimension) and uniform (assayed with the lacunarity) throughout the larval stages.
To understand how the class IV neuron maintains its morphology during larval development, we constructed a mathematical model based on random branching and self-avoidance.
We found that if the branching rate is uniform in space and time and that if all contacting branches are deleted, we can reproduce the branch length distribution, mesh size and density of the class IV dendrites throughout the larval stages. 
Thus, a simple set of statistical rules can generate and maintain a complex branching morphology during growth.
\end{abstract}

%\contributor{Submitted to Proceedings of the National Academy of Sciencesof the United States of America}
%\keywords{Drosophila melanogaster | class IV da neuron | morphology | dendrite}

In our brains, billions of neurons interact with each other to build a nervous system of unparalleled complexity and computational power. 
Neurons have dendrites, which are branched structures that receive synaptic or sensory inputs, and an axon, which send outputs to other neurons. 
The shape or morphology of individual neurons sets the number and types of interactions that a neuron can have and provides the structural basis of neuronal computation \cite{Wassle:1991lq, Hausser:2000pi, Attwell:2001bq, Shepherd:2005fp, Spruston:2008eu, Wen:2008fv}.

Since many organisms continue to enlarge after the establishment of the body plan, it is critical for axons and dendrites to maintain their morphology as they grow.
For example, interneurons of the grasshoper \cite{Bentley:1981kq}, motor neurons in moths \cite{Truman:1988rw} and mice \cite{Li:2005vn} grow drastically in size yet maintain connections to their target cells. 
Futhermore, dendrites in the perpherial nervous system, like those of gold fish retinal ganglion cells \cite{Hitchcock:1987sf}, and dendritic arborization (da) sensory neurons of the fly larva \cite{Parrish:2009pb}, which are the subject of this work, grow to continually maintain coverage of their receptor fields. 
In this paper, we investigate the growth rules that are required to maintain the correct branching morphology as a dendrite grows.

The da sensory neurons of the fly larva are a model system for studying dendritic arborization \cite{Jan:2010uk, Zipursky:2013jk, Dong:2015sf}. 
These dendrites innervate the extracellular matrix, which lies between the outer cuticle and the inner epidermal cell layer \cite{Bodmer1987}. 
They tile the surface on the fly larva in a highly stereotyped manner and have four distinct morphological classes \cite{Grueber:2002sf} (Fig. \ref{fig:classes} A).
Since it is easy to identify and image individual da neurons, these neurons have proven to be a powerful model system for studying dendrite morphology \cite{Jan:2010uk, Zipursky:2013jk}.
In this paper, we address the question of how the morphology of the class IV da neurons (Fig. \ref{fig:classes} B) is maintained during the larval stages.

The class IV da neuron has highly branched dendrites \cite{Grueber:2002sf}, which detect potentially harmful stimuli, such as the ovipositor barb of parasitic wasps \cite{Hwang:2007qq, Robertson:2013dz}. 
The dendrites of the class IV neuron begin morphogenesis during late embryogenesis $\sim 16\,\text{hrs}$ After Egg Lay (AEL).
By the time the larva hatches ($\sim 22\,\text{hrs}$ AEL at $\SI{25}{\degreeCelsius}$), the class IV dendrites nearly cover its surface. 
The dendrites then continue to expand and branch as the larva grows ($22-126\,\text{hrs}$ AEL), so that the neuron maintains its coverage of the larval surface \cite{Parrish:2009pb}.

In this work, we are seeking the growth rules that allow class IV dendrites to maintain their dense coverage of the larval surface.
To this end, we have used a novel measure, the hitting probability, that quantifies the mesh size and two well-known measures of branching morphology: the fractal dimension \cite{Mandelbrot:1967yg} and lacunarity \cite{Mandelbrot:1982pd,Allain:1991cq} (see \Term). 
We show that these measures remain largely invariant over larval stages, despite a several fold increase in larval length. 
Furthermore, we demonstrate that a model with simple rules for branching and self-avoidance can capture essential features of the establishment and maintenance of the dendrite's morphology.

\section{Experimental Results}
To characterize the morphology of fully-developed class IV dendrites, we imaged larvae expressing Cd4-tdGFP under the ppk promoter (\textit{ppk-cd4-tdGFP}) during the third instar stage (Fig. \ref{fig:classes}) using a laser-scanning confocal microscope (See Material and Methods for details). 
Using NeuronStudio \cite{Wearne:2005tg} and Fiji we traced the branches of the dendrites to produce skeletons. 
These skeletons were then analyzed to obtain the the mesh size, density and uniformity of class IV dendrites using parameters defined in the next section.
\subsection*{\Term}
Here we include simple definitions of the relevant morphometrics to aid comprehension.
\vspace{-.2cm}
\begin{description}
\item[Hitting Probabiltiy] $H(B)$: The probability that a box of size $B$ hits the dendrite.
\item[Mesh Size] $\hittingscale$: The length at which $50\%$ of all boxes hit the neuron.
\item[Fractal Dimension] $\df$: A measure of the space-fillingness of a shape. For a completely filled box $\df = 2$, for a straight line $\df = 1$, for branched shapes $1\le\df\le 2$.
\item[Lacunarity] $\lac(B)$: A measure of density fluctuations as a function of length scale $B$. 
\item[Lacunarity Length] $\lacscale$: The length at which $\lac(B=\lacscale) = 0.25$, i.e. the length at which the neuron is uniform. The larger $\lacscale$, the more variable the density of the neuron.
\item[Radius of Gyration] $\Rg$: A length scale that measures how spread out a shape is from its center. The larger $\Rg$ the more spread out the neuron.
\item[Persistence Length] $\perL$: The characteristic length at which a branch bends.
\end{description}
For mathematical definitions see \textbf{Appendix}.
\subsection{Class IV dendrites have a small mesh size}
\begin{figure*}[h!]
		\centering
		\includegraphics[width = 1\textwidth]{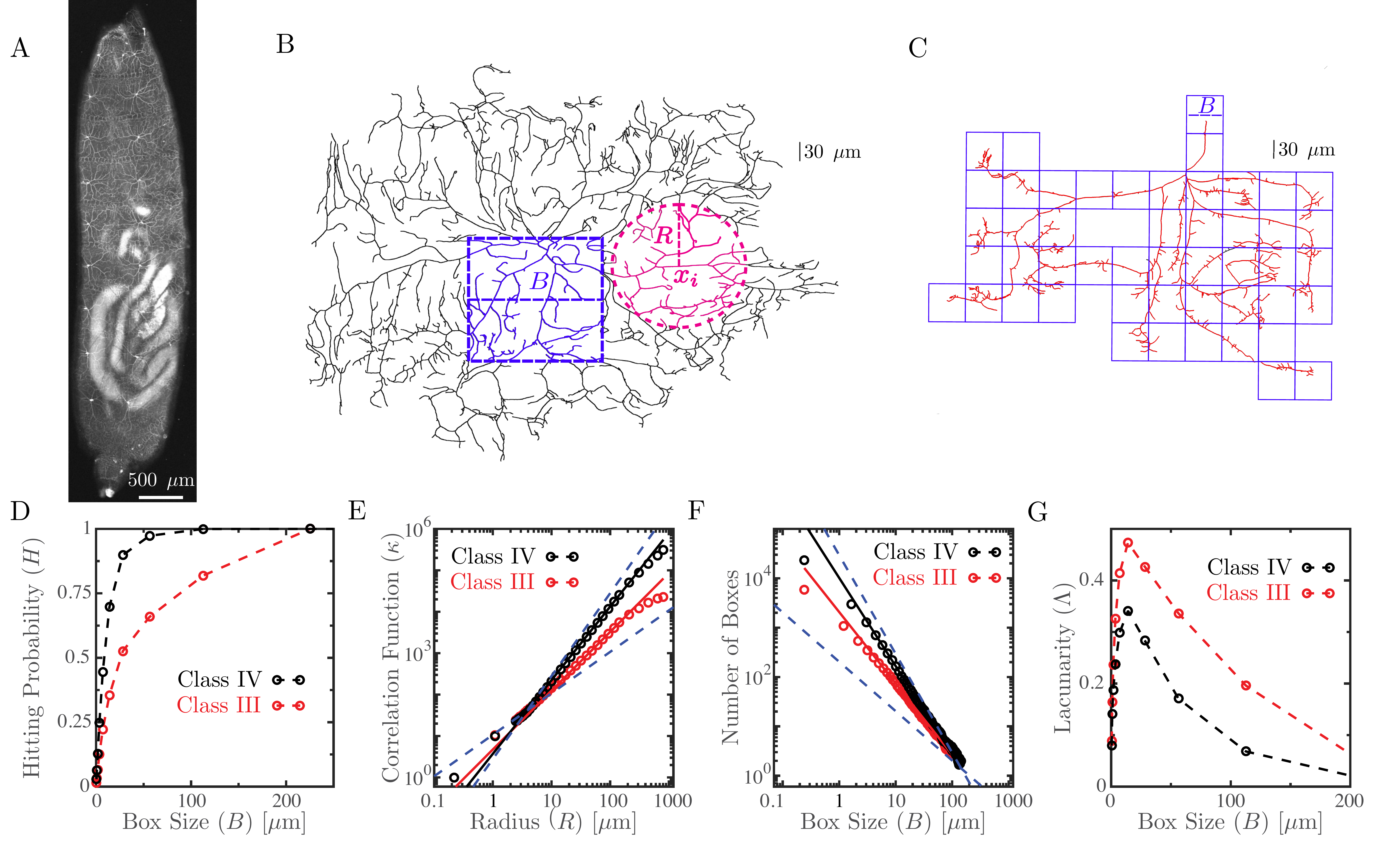}
		\caption[]{\textbf{Morphometrics of class IV and class III dendrites.}
		\textbf{(A)} $3^\text{rd}$ instar larval expressing GFP-tagged membrane protein (\textit{ppk-cd4-tdGFP}) in class IV neurons. 
		\textbf{(B)} The skeleton of a class IV neuron from a third instar larva with an example of a box used to calculate the hitting probability (blue) and a circle used to calculate the correlation dimension $\dc$ (magenta). %The hitting probability $H(B)$ is computed by determining the probability that a box of size $B$ has at least one pixel of neuron in it (Eq. \ref{eq:Hit_int}). To calculate the correlation dimension we measure the average number of pixels $\kappa$ in a circle with radius $R$ (Eq. \ref{eq:corrfun}).
		\textbf{(C)} The skeleton of a class III neuron from a third instar larva with an example of a set of boxes used to compute the box dimension $\db$ and lacunarity function $\lac$. See \textbf{\Term} and \textbf{Appendix} for details %The box dimension is determined by counting the number of boxes of size $B$ need to cover the neuron. The lacunarity function is the square of the coefficent of variation in the mass per box as a function of the box size $B$ (Eq. \ref{eq:lacfun}).
		\textbf{(D)} Example hitting probability $H$ versus box size $B$.
		\textbf{(E)} The correlation function $\kappa$ versus the radius $R$. Fits used to determine the correlation dimension $\dc$ plotted in solid lines. See \textbf{Appendix} for details of fits. The curves for $\dc = 1$ and $\dc = 2$ are plotted in blue for reference.		
		\textbf{(F)} The number of boxes needed to cover a neuron versus the box size $B$. Fits used to determine $-\db$ are plotted in solid lines. See \textbf{Appendix} for details of fits. The curves for $\db = 1$ and $\db = 2$ are plotted in blue for reference. 
		\textbf{(G)} Lacunarity function $\lac$ versus box size $B$ for a class IV neuron (black) and a class III neuron (red). 
		\label{fig:classes}}	
\end{figure*}

%To understand the relationship between the coverage and length scale for a given neuron, we developed a novel measure called the hitting probability $H(B)$. 
To characterize the mesh size of the dendrites, we developed a novel measure called the hitting probability $H(B)$.
$H(B)$ measures the probability $H$ that a randomly placed box of size $B$ hits the dendrite (see Appendix for details).
%$H$ generalizes the previously used coverage index by allowing us to quantify the effect of box size on coverage.
%Previous work \cite{Parrish:2009pb} used the coverage index which measures the fraction of boxes $1/12$ the size of the neuron, in a grid, that contain neuron. 
The hitting probability generalizes an earlier metric called the coverage index  \cite{Parrish:2009pb} by allowing for any box location and any box size. 
%Intuitively, the smaller the box needed to achieve a high hitting probability the better the coverage. 
%If small boxes have a high hitting probability, the mesh size is small.
A typical hitting probability curve of a neuron (Fig. \ref{fig:classes} D) $H(B)$ increases monotonically with $B$, eventually reaching $H=1$ as $B$ approaches the size of the neuron.

%The major advantage of the hitting probability over the previously used measures of coverage is that it allows us to understand the role of the spatial scale. 
%In particular, we can define the scale $\hittingscale$ at which $H(B=\hittingscale) = 0.5$.
We define the characteristic mesh size $\hittingscale$ as the box size at which half of all boxes hit the dendrite.
In other words, $\hittingscale$ is the maximum size of a stimulus that would go untouched, or undetected, on average, by the neuron. $\hittingscale$ is similar to the mesh size in a cross-linked polymer network\cite{Doi:1988vn}.

We found that $\hittingscale = 8.4 \pm 0.5\mm\, (\textrm{mean}\pm\textrm{SD},\,n = 14\,\textrm{neurons})$ for the mature dendrites of the class IV neuron.
Thus, the mesh size is approximately equal to the diameter of the ovipositor barb of wasps that lay eggs in \textit{Drosophila} larva ($\sim 10 \mm$, \cite{Hwang:2007qq}).
This indicates that the class IV dendrite has a high chance of detecting a wasp attack.
Furthermore, the mesh size is small compared to the overall size of the neuron ($\sim 500\mm$) and is similar to the mean branch length (see below).

\subsection{Class IV dendrites are dense and uniform}
%The fractal dimension is the scaling exponent of the density $\rho$. 
%For planar shapes like the class IV, the density is the mass $M$ contained is a circle $R$ divided by the area of the circle, i.e. $\rho \propto M/R^2$. 
%For a filled circle (or box) $M \propto R^2$, so $\rho$ is a constant, regardless of the size of the circle.
%But for shapes with holes and branches, this is not true. 
%In these cases, the density will be $\rho \propto R^{\df-2}$, where the scaling exponent $\df$ is called the fractal dimension \cite{Mandelbrot:1967yg}.
To understand the morphological basis underlying $\hittingscale$ we quantified the density and uniformity of the class IV dendrites during the third instar stage.
The fractal dimension $\df$ is a commonly used measure of how a branched structure fills space. 
A solid square, for example, has a $\df = 2$, while a straight line has a $\df = 1$ (See Appendix for mathematical definitions of $\df$). 
We found that the fractal dimension of class IV dendrites was $\df =1.80\pm 0.04\,(\textrm{mean}\pm \textrm{SD},\,n=14)$, indicating that the dendrites are dense and space filling. 

To establish a small mesh, a dendrite needs to not only be space filling (i.e. $\df\sim2$), but uniformly so. 
To measure the uniformity of the dendritic arbor, we used the lacunarity function $\lac(B)$.
This measures the density variation as a of function length scale $B$. 
%$\lac(B)$, is the variance in the number of neuron skeleton pixels (i.e. mass) in a box of size $B$ divided by the mean number of pixels, as a function of the box size (see \Term). 
To compare the lacunarity of different cells we calculated the length $\lacscale$ ($\lac(B = \lacscale) = 0.25$ see \Term). 
The smaller $\lacscale$, the more uniform the dendritic density.
We found that $\lacscale = 32.6 \pm 16.8\mm \,(\textrm{mean}\pm\textrm{SD},\,n = 14)$ for third instar larva.
In other words, on spatial scales larger than $33\mm$ the density of the neuron is uniform, whereas below $33\mm$ it is variable.
While the $\lacscale$ is larger than the mesh size $\hittingscale$, it is much smaller than the dendrite size, indicating that the coverage is uniform and arbors can be considered homogeneous.
In summary (Tab. \ref{tab:cl3vcl4}), mature class IV neurons have dense (large $\df$) and uniform (small $\lacscale$) dendritic arbors, which is consistent with a small mesh size (small $\hittingscale$). 
%They also have a qualitatively different morphology, which may reflect the very different mechanosensory functions. 

\subsection{Class IV dendrites have a tighter mesh, are denser and more uniform than class III dendrites}
To assess the ability of these measures to quantify dendritic morphology, we also imaged class III cells (Fig. \ref{fig:classes} C). 
Class III cells are gentle touch sensors and use a different set of mechanotransduction channels \cite{Yan:2013fr}.
The class III dendrites (Fig. \ref{fig:classes} C) are substantially less branched and have smaller branches, on average, than the class IV dendrites (Tab \ref{tab:cl3vcl4}).
We find that $H(B)$ is much smaller for class III neurons than for class IV neurons (Fig. \ref{fig:classes} D) for most box sizes. 
%This smaller $H$ across most scales demonstrates, quantitatively, that the class III neuron is worse at covering the larva surface than the class IV neuron.
Consequently, we find that the mesh size $\hittingscale$ is much larger in class III neurons ($\hittingscale \sim 24.7\mm$) than in class IV neurons ($\hittingscale \sim 8.4\mm$)  (Tab. \ref{tab:cl3vcl4}). 
In other words, class III dendrites have larger gaps in coverage than class IV dendrites.
%We would expect that this would mean that class III neurons are also sparser and less uniform than class IV dendrites.
\begin{table*}[ht]
	\centering
	\caption{Properties of class IV and class III neurons. All numbers are mean $\pm$ SD.\label{tab:cl3vcl4}}
	\begin{tabular*}{\hsize}{@{\extracolsep{\fill}}lcr}
		 &class IV$\,(\text{n}=14)$&class III$\,(\text{n}=8)$\cr
		\hline
		Mean branch length, $\meanL$&$12.54\pm0.04\mm$&$7.49\pm0.8\mm$\cr
		Mesh size, $\hittingscale$&$8.37\pm1.80\mm$ &$22.7\pm4.8\mm$\cr
		Fractal dimension (correlation method), $d_\text{c}$&$1.80\pm0.04$&$1.42\pm0.03$\cr
		Fractal dimension (box method), $d_\text{b}$&$1.79\pm0.04$&$1.43\pm0.04$\cr
		Lacunarity length, $\lacscale$&$32.6\pm16.8\mm$ &$130.5\pm31.2\mm$\cr
		\hline
	\end{tabular*}
\end{table*}

We find that class III dendrites have a fractal dimension of $\df \sim 1.42$ (Tab \ref{tab:cl3vcl4}), which is substantially smaller than the class IV neuron. 
In other words, class III dendrites are sparser, at all scales, than class IV dendrites (Fig. \ref{fig:classes} E and F).
We find that the lacunarity $\lac$ decays much slower (with $B$) for class III dendrites compared to class IV dendrites (Fig. \ref{fig:classes} G). 
Consequently $\lacscale$ is much larger (Tab. \ref{tab:cl3vcl4}) for class III dendrites than for class IV neurons (Tab. \ref{tab:cl3vcl4}) showing that they are less uniform than class IV neurons.
These results (Tab. \ref{tab:cl3vcl4}), show that class III neurons are sparser (small $\df$) and less uniform (large $\lacscale$) than class IV dendrites, which is consistent with having larger mesh size (large $\hittingscale$). 
These differences likely reflect the different mechanoreceptor functions of class III and class IV neurons (see Discussion).
\begin{figure*}[ht]
	\centering
	\includegraphics[width=1\textwidth]{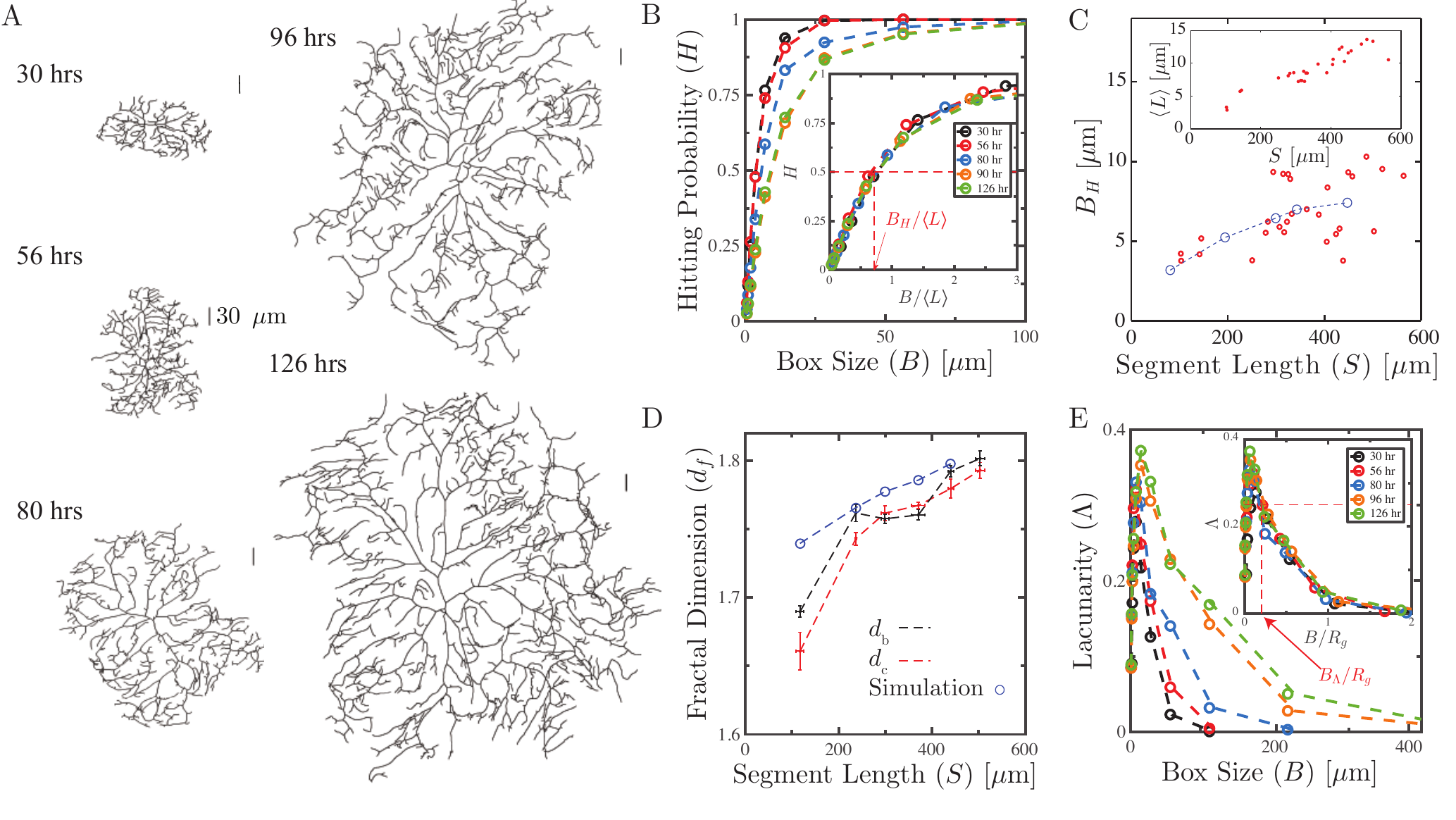}
	\caption{\textbf{Morphometrics of class IV neurons during larval growth.}
	\textbf{(A)}Examples of class IV neurons during larval stages. All scale bars 30$\mm$, all time stamps hours After Egg Lay. 
	\textbf{(B)} The hitting probability $H$ is plotted versus the box size $B$ for the $5$ neurons shown in Fig. \ref{fig:Morphvtime} A. In the inset, we plot $H$ versus $B/\meanL$. We define the mesh size $\hittingscale$ such that $H(B = \hittingscale) = 0.5$. $\hittingscale \sim 0.72\meanL$ throughout the larval stages.
	\textbf{(C)} $\hittingscale$ is plotted versus larva body segment length $S$ (red $28$ cells). For simulated neurons, $\bfreq = 0.2\minute$ and $\retL = 10^{2}\mm$ (blue).
	In the inset, we have plotted the mean branch length of the class IV dendrites versus the larva body segment length $S$. 
	\textbf{(D)} The fractal dimension $\df$ is plotted against $S$. We have binned the data by body segment length $S$ with bin widths of $77\mm$. Simulation parameters are same as before.
	\textbf{(E)} The lacunarity $\lac$ is plotted against box size $B$ for the five neurons in Fig. \ref{fig:Morphvtime}. In the inset, we plot $\lac$ versus $B/\Rg$. We define the lacunarity length $\lacscale$ as the box size at which $\lac(B = \lacscale) = 0.25$. 
	\label{fig:Morphvtime}}
\end{figure*}

\subsection{Class IV dendrites maintain a tight mesh, high density and uniformity throughout larval stages}
To determine how the morphology of class IV dendrites changes with time, we imaged and skeletonized the class IV dendrites, as before, from early first instar ($30\,\text{hrs AEL}$) to wandering third instar ($126\,\text{hrs AEL}$) (Fig. \ref{fig:Morphvtime} A). 
We used the larval body segment length as a proxy for time since each data point comes from a unique larva, and $S$ increases linearly with time \cite{Ashburner:2005qf} (SI).

%Analysis of class IV dendrites in larvae of different ages (Fig. \ref{fig:Morphvtime} A) showed that the hitting probability decreases with time (Fig. \ref{fig:Morphvtime} B), i.e. $\hittingscale$ increases during the development (Fig. \ref{fig:Morphvtime} C). 
%This change in $H(B)$ is likely related to the increase in the mean branch length $\meanL$ (Fig. \ref{fig:Morphvtime} C inset) since the hitting probability curves $H(B/\meanL)$ for all ages follows the same curve (Fig. \ref{fig:Morphvtime} B inset), i.e. 
The mesh size $\hittingscale$ increased modestly from $4.4\pm0.3\mm $ (1st instar) to $8.5\pm1.4\mm$ (third instar) (Fig. \ref{fig:Morphvtime} C). 
This two-fold increase is less than the six-fold growth in larval body segment size and indicates that the mesh size remains small during development. 
Interestingly, the ratio between $\hittingscale$ and the mean branch length $\meanL$ is nearly constant during development showing that the shape of the arbor has remarkable conservation during development
%Though the hitting probability changes during the larval stage, $\hittingscale$ only increases by $\sim 2$ folds and was always below $10\mm$ (Fig. \ref{fig:classes} C), i.e. the mesh size is always small compared to the body segment size during the larval stages.

The morphologies of the class IV neurons remained dense and uniform during the larval stages of development.
We found that the fractal dimension (Fig. \ref{fig:Morphvtime} D) in the just hatched larvae ($\sim30$hrs AEL) was $\df \sim 1.7$ and increased to $1.75$ within 24 hours, eventually rising to about $1.8$ during the next four days.
The lacunarity slightly increased with time (Fig. \ref{fig:Morphvtime} E). By rescaling $B$ by the radius of gyration $\Rg$ (i.e. overall neuronal size see Eq. \ref{eq:rg}), we found that $\lac$ follows the same curve at all developmental times, i.e., collapses when scaled by $\Rg$ (Fig. \ref{fig:Morphvtime} D inset).
The nearly constant fractal dimension and the collapse of the lacunarity curves indicate that the morphological pattern of the class IV neurons is mostly established during embryogenesis.% and maintains the same relative shape.

\section{Dynamic Model of Class IV Development}
\begin{figure*}[ht]
	\centering
	\includegraphics[width = 1 \textwidth]{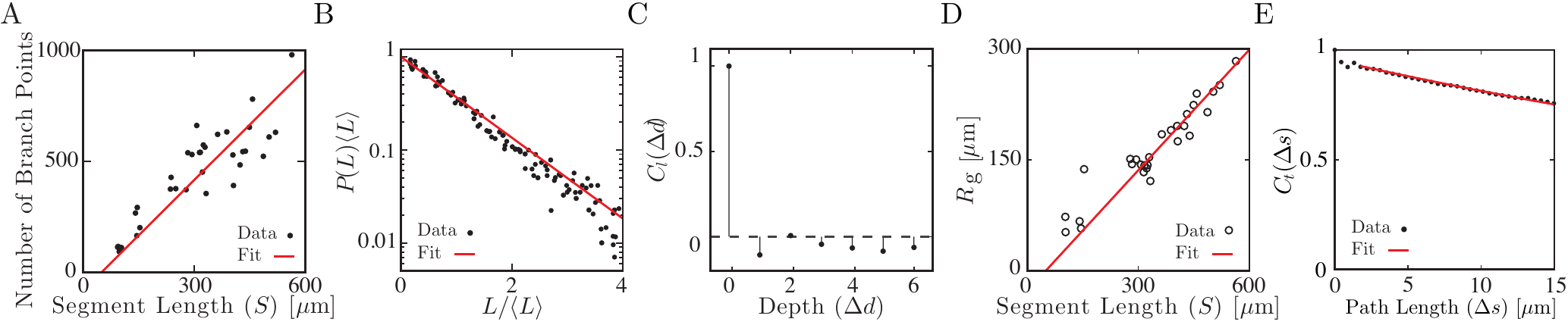}
	\caption{\textbf{Branching Rules}
	\textbf{(A)} Number of branches versus larval body segment length $S$ for class IV dendrites at different larval stages. The slope of the fit is $1.64\,\text{branches}\mm^{-1}$ for the line passing through the $S$ axis at $50\mm$, the segment length at the onset of dendritogenesis.
		\textbf{(B)} The probability density $P(L)$ that a branch of a dendrite (at a particular time) has length $L$. The probability density is rescaled by multiplying by $\meanL$ and the branch length is rescaled by dividing by $\meanL$ (of the particular dendrite). The superposition of the data indicates that the branch lengths are well described be an exponential distribution (red line) at all larval stages.
	\textbf{(C)} The branch length autocorrelation $C_l$ (Eq. \ref{eg:blcorr}) vs depth difference $\Delta d$ averaged over all branches.
The near-zero values for $\Delta d \ge 1$ imply that branch length is independent of depth.
	\textbf{(D)} The radius of gyration $\Rg$ (Eq. \ref{eq:rg}) of $28$ class IV neurons versus larval body segment length $S$. The slope of the fit is $0.39$, where the $S-$intercept is set such that $\Rg = 0$ at the onset of class IV dendrite morphogenesis.
	\textbf{(E)} The tangent vector autocorrelation function $C_{t}$ (Eq. \ref{eq:persis}), averaged over all branches, versus the path length lag $\Delta s$ (Eq. \ref{eq:persis}). The red line is an exponential fit to the data $C_t(\Delta s) = e^{\Delta s/\beta}$, where $\beta = 44.8 \pm 1.5\mm$ is the persistence length.
	\label{fig:modelinputs}}
	\hspace{-1cm}
\end{figure*}

The maintenance of a small mesh size throughout larval stages raises the question: how can the class IV dendrites achieve this, despite the six-fold increase in the larval segment size?
To answer this question, we developed a mathematical model of class IV dendrite morphogenesis during the larval stages.
Our model consists of three rules that determine 1) branch creation, 2) the direction and speed that this new branch grows, and 3) how branches avoid crossing the other branches (self-avoidance).  
%Our model consists of three rules that determine 1) branch creation, 2) branch tip elongation and 3) branch self-avoidance.
%This model was focused on exploring the cell biological basis of these controlling mechanisms and was based on four observations: 1) the overall size of the neuron increases, 2) the branches are relatively straight 3) the branches do not cross and 4) the number of branches increases.

\subsection{Branch Creation} 
We found that the number of branch points increases with time during larval stages (Fig. \ref{fig:modelinputs}A), and this increase was well described by a linear function with a net branch creation rate $\bfreq_\text{net} \sim 0.1\,\text{branch points}\minute$ (Tab. \ref{tab:modelparam}). 
This observation led us to model branch creation as a time invariant process with a branching frequency $\bfreq$.
Since $\bfreq_\text{net}$ can include the removal of branches (see below), it is a lower bound on the branching creation frequency $\bfreq$.

We also measured the branch length distribution, and found that it was well described be an exponential distribution (Fig. \ref{fig:modelinputs} B). 
Furthermore, we found that that the branch length was independent of the branch depth, defined as the number of branch points between a branch and the soma along the shortest path (Fig. \ref{fig:modelinputs} C). 
Motivated by these observations, we modeled branch creation as a random process that was uniform along the neuron and constant in time.

\subsection{Tip Elongation} 
Since the growth of class IV dendrites occurs at the branch tips \cite{Matthews:2007kq,Parrish:2009pb}, we modeled neuron growth as branch tip elongation. 
We measured the overall size of class IV dendrites using the radius of gyration $\Rg$ (Eq. \ref{eq:rg}). 
$\Rg$ measures spread of neuron mass from its center. 
We found that $\Rg$ increases linearly during development (Fig.  \ref{fig:modelinputs} D).
Therefore, we assume that the tip elongation rate $\growthspeed$ is constant in time and $\growthspeed \propto \Vg$, where $\Vg$ is the growth speed of the class IV dendrite.
The assumption of a constant growth speed assumes that the simulation time scale is much larger than the fluctuation times scale. %at the time scales considered we can ignore fluctuations in the tip elongation rate.
We estimated $\Vg$ from the change in dendrite size during development (Fig. \ref{fig:modelinputs} D), and found $\Vg\sim 0.04 \mm \min^{-1}$ (Tab \ref{tab:modelparam}). 

%To measure $\Vg$, we fit a line to Fig. \ref{fig:modelinputs} D, where we set the $S-$intercept $= 50\mm$ so that $\Rg = 0\mm$ at the onset of class IV dendrite morphogenesis. 
%Then using the mean larval growth rate we convert the slope of the fit to a growth speed $\Vg\sim 0.04 \mm \min^{-1}$ (Tab \ref{tab:modelparam}). 
%To complete our tip elongation rule, we need to determine the direction that the branch tips grows in.

To determine the direction of branch growth, we measured how much the path of a branch changes as a function of path length by using the persistence length $\perL$; the distance over which the orientation of the direction of growth of a branch changes (see Appendix for mathematical details). 
We find that $\perL = 45\pm2\mm$ (Tab. \ref{tab:modelparam}, Fig. \ref{fig:modelinputs} E), which is much greater than the mean branch length ($\meanL \sim 12.5$ Tab. \ref{tab:cl3vcl4}), indicating that branches tend to be straight.

%It is important to note that if there exists un-branching in class IV dendrites, then sudden changes in the path of a branch can be created, which would reduce the measured $\perL$. 
%These unexpected turns can occur if the primary daughter branch, defined as the daughter whose direction of growth is closest to the mothers, where to un-branch.
%Then the path would appear to have a sudden turn at the location of the secondary daughter branch, which would have grown, on average, perpendicular to the primary daughter.
%As such the measured $\perL$ is a lower bound, and in our simulations we use a slightly larger $\perL$.

\subsection{Self-Avoidance} 
Previous work has shown that the branches of the class IV neuron do not cross each other. 
Furthermore, it is known that this self-avoidance is contact mediated and it has been proposed that branches retract after contact \cite{Matthews:2007kq,Soba:2007yu,Hutchinson:2014ul,Matthews:2011vn}.
Therefore, we modeled self-avoidance by having growing branches retract at a constant speed $\growthspeed$ (same as elongation rate), if they contact an existing branch.
For simplicity, the retraction length was assumed to be exponentially distributed with a mean retraction length $\retL$.
\begin{table*}[ht]
\centering
\caption{Parameters in the model. All errors are SD (n=28 neurons).\label{tab:modelparam}}
	\begin{tabular*}{\hsize}{@{\extracolsep{\fill}}lcr}
		Parameter & Measured Value & Simulation Value \cr 
		\hline 
		Branching frequency \comment{$\bfreq_\textrm{net}$ and $\bfreq$)} & $0.12\pm 0.03\min^{-1}$ ($\bfreq_\textrm{net})$ & $0.01\,-\,2\min^{-1}$($\bfreq$) \cr
		Tip elongation rate, $v$ & $0.04\pm0.02\mm\cdot\min^{-1}$ & $0.08\mm\cdot\min^{-1}$ \cr
		Persistence length, $\perL$ & $44.8\pm1.5\mm$ & $70\mm$ \cr
		Retraction length, $\retL$ & not measured & $0-1000\mm$ \cr
		\hline 
	\end{tabular*} 
\end{table*}
\subsection{Model Implementation}
Using these rules, we arrived at a four parameter model; $\bfreq$ the branch creation rate, $\growthspeed$ the growth speed, $\perL$ persistence length and $\retL$ to retraction scale. 
We implemented our model on a hexagonal lattice, with lattice spacing $\epsilon = 0.4\mm$. 
We chose to use a lattice model to facilitate contact detection as the lattice spacing was set to be equal to the thickness of the branches.
Since we implemented the rules on a hexagonal lattice, the possible branching angles are limited. 
Therefore we ignored the role of branching angles on morphology.
The lattice spacing and the growth speed act as scale factors, i.e. setting the overall size of the dendrite but not changing the shape.
Therefore, we chose $v$ such that the size of a simulated dendrite matched that of a real one, which left us with three free parameters. 
However, we found that as long as $\perL$ is large, i.e., as long as the branches are straight, $\perL$ has little effect on the morphology (SI).
Therefore, we focused on how branching frequency $\bfreq$ and the contact-based retraction scale $\retL$ control the morphology of dendrites.

\subsection{Model Results}
\begin{figure*}[ht]
\centering
	\includegraphics[width= 1 \textwidth]{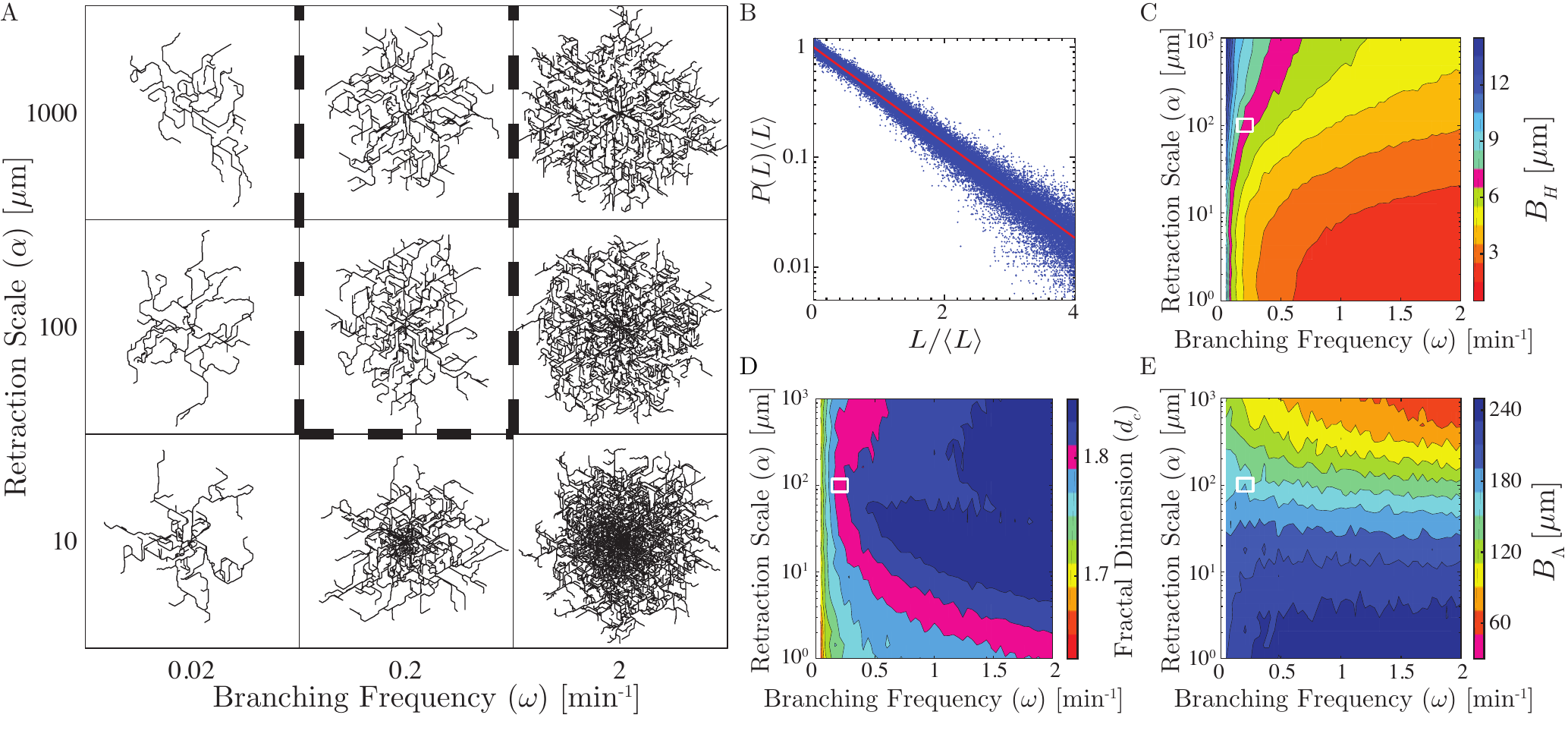}
	\caption{\textbf{Simulations of dendritic growth}
	\textbf{(A)} Nine representative simulations of dendrties for a range of branching frequencies $\bfreq$ and retraction scales $\retL$. 
		For all simulations, the persistence length was $\perL = 70\,\mm$ and simulation time was $\totsimtime = 100\,\text{hrs}$.
		For $\retL\gg 1\,\mm$ and $\bfreq \sim 0.2\text{min}^{-1}$ (dashed box) we have subjective agreement with the morphology of real neurons.
		%This region is highlighted by dashed lines.
	\textbf{(B)} The branch length distribution for simulated neurons in the dashed black box in \textbf{(A)} is approximately exponentially distributed as observed.
	\textbf{(C)} The mesh size $\hittingscale$ in a contour plot versus the branching frequency $\bfreq$ and retraction scale $\retL$.
Note that the retraction scale $\retL$ is plotted in a log scale. 
In pink we have highlighted the observed value of $\hittingscale = 7\mm$ (see Tab. \ref{tab:cl3vcl4}). The white box indicates the values used for the time series plotted in Fig. \ref{fig:Morphvtime} C and D ($\bfreq = 0.2\minute$ and $\retL = 10^{2}\mm$).
	\textbf{(D)} The fractal dimension $\df$ in a contour plot versus the branching frequency $\bfreq$ and the retraction scale $\retL$.
%Also, we notice that $\df$ saturates for $\bfreq>0.5\minute$.
In pink we have highlighted the physiological value of $\df = 1.8$ (see Tab. \ref{tab:cl3vcl4}) 
	\textbf{(E)} The lacunarity scale $\lacscale$ in a contour plot versus the branching frequency $\bfreq$ and the retraction scale $\retL$.
We highlighted in pink the observed value $\lacscale = 33\mm$ (see Tab. \ref{tab:cl3vcl4}). We find that this value is not found in the range of values used in our simulations. %Indicating that the simulated dendrites have a larger spatial variation in density than the real neurons.
\label{fig:modeloutputs}}
\end{figure*}

We generated simulated neurons for a range of branching frequencies $\bfreq$ and retraction scales $\retL$ (Fig. \ref{fig:modeloutputs} A) and analyzed their shape.
Importantly, we found that the distribution of branch lengths was exponential (Fig. \ref{fig:modeloutputs} B) and the branch lengths were uncorrelated with branch depth, for a limited range of parameters (dashed box in Fig. \ref{fig:modeloutputs} A), which agrees with the experimental observations (Fig. \ref{fig:modelinputs} B and C).

To test whether our model provides a good description of the morphology, we measured the mesh size, fractal dimension and lacunarity of the simulated dendrites.
The mesh size $\hittingscale$ decreased with increasing $\bfreq$ and decreasing $\retL$ (Fig. \ref{fig:modeloutputs} C) and appeared to saturate as $\bfreq$ increases.  
There was a small region of $\bfreq-\retL$ space where there was quantitative agreement between the simulated and observed values of $\hittingscale$ (Fig. \ref{fig:modeloutputs} C, pink).  
We found, that $\df$ increased monotonically with $\bfreq$ and saturated for large values of $\bfreq$. 
As for $\hittingscale$, we found a narrow band of $\bfreq-\retL$ space where we have a quantitative agreement between the model and the class IV neuron for $\df$ (Fig. \ref{fig:modeloutputs} D, pink).
Crucially, there was a small region of $\bfreq-\retL$ space where the model recapitulated both $\hittingscale$ and $\df$ (Fig. \ref{fig:modeloutputs} C and D, white box). 
Taking values from this small region ($\bfreq = 0.2\,\text{min}^{-1}$ and $\retL = 10^2\mm$), we recapitulated the third instar and even found agreement throughout most of the larval development (Fig. \ref{fig:Morphvtime} B and \ref{fig:Morphvtime} C).
Thus, the model recapitulates both the mesh size and fractal dimension throughout development.
It did not recapitulate the lacunarity, which for the parameter values $\bfreq=0.2\,\text{min}^{-1}$ and $\retL = 10^2\mm$ was larger than the observed values.
In other words, the model arbors are not as uniform as the observed ones (see Discussion). 
In summary, this model recapitulates most, but not all aspects of the dendritic morphology of class IV neurons.

%We then proceeded to measure the lacunarity (Fig. \ref{fig:modeloutputs} B). 
%The most striking feature of the lacunarity is the horizontal stratification of $\lacscale$, i.e. that there is little to no change in $\lacscale$ with $\bfreq$. 
%We think that the stratification is due to the dense core and sparse periphery that we noticed in neurons with small $\retL$ (bottom right Fig \ref{fig:modeloutputs} A). 
%As we show in the SI, the dense core leads to an increase in $\lacscale$. 
%We notice that as $\retL$ increases $\lacscale$ decreases, implying that deletion of branches makes the dendrites more uniform.
%Unlike with $\hittingscale$ and $\df$, there is little to no overlap between the physiological values of $\lacscale$, $\hittingscale$ and $\df$ (pink bands in Fig. \ref{fig:modeloutputs} B, C and D). 

%These results showed that branching positively contributes to the neuronal complexity by promoting the dendritic growth while the contact-based retraction provides precise control by negatively controlling net growth and preventing over-filling of the receptor field. 
%These two process together shape the morphological pattern of the class IV dendrite and set the dendrite mesh size.
\section{Discussion}
%Our main experimental finding is that as the dendritic arbors of class IV cell grow during larval development, the morphology, as characterized by several parameters including branch length, mesh size, fractal dimension and lacunarity remains remarkably constant. 
Our key experimental finding is that the morphology of class IV neurons, as characterized by the branch length, mesh size, fractal dimension, and lacunarity, remarkably remains constant during development.
Indeed, from the early first instar larva ($30$ hours after egg lay) to the late third instar larva ($126$ hours after egg lay), as both the segment size and the number of branches increases approximately six-fold, the mean length, the mesh size, and the lacunarity only increase around two-fold and the fractal dimension is almost unchanged. 
When we normalize the mesh size by the mean branch length, it is virtually unchanged throughout development. 
Thus, as these cells grow, key aspects of their morphology are invariant, even as the overall size the number of branches increases by nearly an order of magnitude. 
%We argue that this invariance may be important for the mechanoreceptor function of these cells, which is to detect highly localized noxious stimuli, necessitating a dense and uniform covering of the body surface (see below).

As a first step toward probing the mechanism underlying this geometric invariance, we developed a simple computational model for branched morphogenesis. 
The model assumes that the branching rate is constant over development (consistent with the observed linear increase in the number of branches), that the rate was independent of position and that growing tips retract random, exponentially distributed distance after contacting other branches. 
The model reproduced many of the key features of the growth of class IV cells including branch lengths, mesh sizes, and fractal dimensions. 
However, it was unable to capture the lacunarity (the model predicted a higher relative density in the center than was observed, see SI). 
Importantly, the data constrained the values of the branching frequency and mean retraction distance. 
Thus, the model provides a framework for understanding the changes in the morphology of these cells during development. 
   
One of our most striking experimental and theoretical findings was that the branching frequency in class IV dendrites was independent of total dendrite length. 
Naively, we might have expected that the mean number of branches added per unit time would increase with total dendrite length, as the longer the dendrites, the more positions on which branches could form. 
However, this would have led to an exponential increase in the number of branches, rather than the observed linear increase. 
Our modeling shows that even if the retraction length is much larger than the mean branch length, an exponential increase in branch length is still observed, and the distribution of branch lengths deviates from the observed exponential distribution (see SI). 
The constant branching rate suggests that branching is limited by the production of a nucleating factor that is produced at a constant rate.  
Furthermore, our finding that branching is uniform in space implies that the putative nucleation factor would be distributed widely and uniformly throughout the cell. 

We also found that for our model to recapitulate class IV-like morphologies, contact-based retraction needs to lead to complete branch deletion, i.e., the mean retraction distance ($\retL$) is much larger than the mean branch length ($\meanL$). 
By deleting the branches whose tips collide with other branches, gaps of a size similar to the mean branch length are created and maintained. 
Also, dense regions where the gap size is less than or equal to the mean branch length will not increase in density. 
Such overfilling is seen in the simulations for small retraction lengths (Fig. \ref{fig:modeloutputs} A). 
Thus, our model constrains both the branching frequency and the retraction distance.

%It is noteworthy that our model slightly overestimates the fractal dimension at the earliest larval stages (Fig \ref{fig:modeloutputs} D). 
%This may indicate that the branching frequency or retraction distance differs between the embryos and the larvae. 
%For example, a lower branching frequency during embryogenesis would support a picture in which the class IV dendrites first grow out to claim their territory (during early development), and then fill in space. 
%Such a picture is supported by our finding that the model predicts a higher-than-observed lacunarity: the density of the simulated dendrites was more spatially variable than the real class IV dendrites. 
%Indeed, the difference arises because the simulated dendrites have a higher density, even when the retraction distance is large. 
%One possible explanation is that neighboring dendrites (or restriction of growth to segment boundaries) force the dendrites to fill in the perimeter, making the overall density more uniform. 
%Again, this supports the idea that the class IV neuron grows out to claim territory, and then fills the space. 

Finally, we note that the mesh size of class IV dendrites, $4 - 8 \mm$, is well suited for detecting highly localized nociceptive stimuli such as punctures by the $10 \mm$ diameter ovipositor barb of parasitic wasps (\cite{Hwang:2007qq}). 
This acuity is maintained throughout development. 
Thus, the small mesh size is consistent with the class IV neuron being a harsh touch sensor. 
Indeed, the theory of contact dynamics predicts that the indentation $h$ of the surface of an elastic body poked by a probe with a cross-sectional radius $R$ (pushing normal to the surface) is $h \propto F/R$, where F is the applied force \cite{Sneddon:1965qf}. 
Therefore, the smaller $R$, the larger $h$ (for a fixed force); the more local the stimuli, the more sensitive to localized forces. 
Thus the small mesh size suggests that the class IV neuron is well adapted to sensing harsh touch throughout larval stages.
In contrast, the class III neuron, a soft touch sensor, would need to capture diffuse stimuli, i.e., the mesh size can be large.
Thus, the morphologies of both class IV and class III neurons are well-suited for their mechanosensitive functions.

\section{Materials and Methods}
	\subsection*{Drosophila Stocks}
	All flies were maintained on standard medium at 23$^o$C. The strain \textit{ppk-cd4-tdGFP} was a kind gift from Dr. Han Chun (Cornell University). 
	
	\subsection*{Imaging and Skeletonization}
	The larvae were mounted in $50\%$ glycerol in PBS between a glass slide and a cover slip. 
	The sample was imaged using a confocal laser scanning microscope (Zeiss, LSM780) with $ 63$x objective. 
	The $600\textrm{x}600\mm$ images were stitched together offline using Fiji and the stitched images were processed using the NeuronStudio \cite{Wearne:2005tg} to obtain the skeleton a one pixel wide tracings of the dendritic arbors.

\appendix
\section{Mathematical Definitions of Morphometrics}
	\subsection*{Radius of gyration of the Neuron}
	The radius of gyration is defined as
	\begin{gather}
	\Rg =\sqrt{\frac{1}{M} \sum_{j = 1}^M \left( {\bf r}_j - {\bf  r}_{\text{m}} \right)^2}, 
	\label{eq:rg}
	\end{gather}
	where $ M$ is the total number of occupied pixels, ${\bf r}_{j}$ is the position of the $ j^{\text{th}}$ occupied pixel and $ {\bf r}_{\text{m}}$ is the mean position of all occupied pixels.
	$\Rg$ measures the standard deviation of the dendrite pixels, i.e., the spread of the imaged neuron from its center.
	
\subsection*{Path Correlation and Persistence Length}	
The deviation of a branch path from a straight line can be quantified using the tangent vector autocorrelation function
\begin{gather}
C_t(\Delta s) \sim \langle\hat{t}(s) \cdot \hat{t}(s+\Delta s)\rangle_s ,
\label{eq:persis}
\end{gather}
where $\hat{t}(s)$ is the tangent vector as a function of the path length $ s$.
$ C_t$ measures the angular change of the $\hat{t}$ as a function of path length, i.e., how bent the branch is. 
If $ C_t = 1$, the path is straight and if $ C_t= 0$, there is a $ 90^o$ turn in the path.

\subsection*{Branch Length Correlation Function}
The branch length autocorrelation function is
	\begin{gather}
	C_l(\Delta d) = \frac{\langle L(d) L(d + \Delta d)\rangle_{d} - \langle L(d)\rangle_{d}^2}{\langle L^2(d)\rangle_{d}-\langle L(d)\rangle_{d}^2},
	\label{eg:blcorr}
	\end{gather}
where $ L(d)$ is the branch length at depth $ d$ and $ \Delta d$ is the depth difference. 
Depth is defined as the number of branch points between the branch and the soma, along the shortest path from the branch to the soma. 
$\langle \ldots \rangle_d$ represents the average over $ d$.
\subsection*{Hitting Probability}
Consider a box with side length $ R$ centered anywhere in the receptor field of the neuron (Fig. \ref{fig:classes} C).
We then ask: `what is the probability $ P_H(b,R)$ of having $ b$ pixels in a box of size $ R$?'. 
Using this probability, we can determine the probability that a box of size $ R$ contains at least $ n$ pixels 
	\begin{gather}
	H_n(R) = \int_n^M P_H(b,R) \mathrm{d}b,
	\label{eq:Hit_int}
	\end{gather}
where $ M$ is the total number of neuron pixels.
	%This is the probability that a stimulus of size $R$ is detected by the neuron, under the assumptions that $n$ pixels are needed to produce a receptor potential that is strong enough to initiate the neuronal action potential, and that the distance to the soma has negligible affect in the firing of the neuron. 
We define the mesh size $ \hittingscale$ such that $ H_1(R=\hittingscale) =0.5$, i.e., the mesh size is the box width such that there is a $ 50\%$ chance that the box contains at least one pixel from the skeleton.
\subsection*{Fractal Dimension}
In this paper, the fractal dimension is measured using two different methods: the correlation dimension (Fig. \ref{fig:classes} E) and box counting (Fig. \ref{fig:classes} F) method. 
In the box counting method, we determine the number of boxes $ N$ of side length $ R$ that are needed to cover the neuron (Fig. \ref{fig:classes} B). 
The number of boxes needed to cover a line of length $ l$ is $ N = \frac{l}{R}$; therefore $ N \propto R^{-1}.$ 
The number of boxes needed to cover a square of side length $ l$ is $ N = \left(\frac{l}{R}\right)^2$; therefore $ N \propto R^{-2}$.
In general, $ N \propto R^{-\db}$, where $\db$ is the box counting measure of the fractal dimension.

In the correlation method \cite{Grassberger:1983pi}, we determine how many pixels are contained within a circle of radius $ R$.
Let each point $ x_i$ on the neuron be the center of a circle of radius $ R$ (Fig. \ref{fig:classes} B).
Then $ N(x_i,R)$ is the number of skeleton pixels in the circle (green pixels in Fig. \ref{fig:classes} C). 
Averaging over all possible centers (i.e. skeleton pixels) $ x_i$, gives 
\begin{gather} 
\kappa(R) = \langle N(x_i,R)\rangle_{x_i}.
\label{eq:corrfun}
\end{gather}
In general, $\kappa(R) \propto R^{\dc}$ where $\dc$ is the correlation measure of the fractal dimension.
		
The relation $ f(R) \propto R^d$ is called a scaling law and is only valid in a finite range of $ R$ (e.g. for small $R$ we approach the scale of one pixel, and for large $R$ we approach the total dimension of the neuron). 
For the neurons the minimum scale is half the mean branch length and the maximum scale is the radius of gyration.\\
\subsection*{Lacunarity} 
Consider the set of boxes of linear dimension $ R$ used in the box counting method (see figure \ref{fig:classes} B). 
Instead of asking how many boxes are needed to cover the shape, we ask `what is the probability $ P_B(b,r)$ of having $ b$ pixels in a box of size $ R$?'. 
$ P_B(b,r)$ differs from $ P_H(b,R)$ since it only considers boxes that have at least one pixel ($ b\geq1$).
	
The moments of $ P_B$ are defined as $$\mu_n(R) = \int_1^M b^n P_B(b,R) \mathrm{d}b, $$ where $ M$ is the total number of skeleton pixels.
This then allows us to look at the coefficient of variation of $ P_B(b,R)$
	\begin{gather}
	CV(R) = \frac{\sigma^2}{\mu_1^2} = \frac{\mu_2 - \mu_1^2}{\mu_1^2},
	\label{eq:lacfun}
	\end{gather}
 where $ \sigma^2$ is the variance of $ P_B(b,R)$. 
$ CV(R)$ is also called the lacunarity function. The more uniform a shape is, the smaller $ CV$ and the more variable the shape, the larger the $ CV$. For example, a uniform shape would have a $ CV\sim 0$. How large $ CV$ needs to be for a shape/neuron to be consider variable is somewhat arbitrary. The more important point is that the larger $ CV$, the larger the variation in the neuron. It also allows us to see how these variations change with length scale. Thus, the lacunarity function measures the variations and assigns them a typical length scale.

\section*{Acknowledgments}
This work was partially supported by the NIH Pioneer Award (Award Number), National Natural Science Foundation of China (NSFC Grant 31671389, to X.L.) and Max-Planck Partner Program (to X.L.). SG was supported by an EMBO Long-Term Fellowship, and OT is supported by the Fonds de Reshershe du Québec - Nature et technologies.
Thanks to Dr. Han Chun (Cornell University) for the fly strains.
%\end{acknowledgments}

\bibliographystyle{unsrt}
\bibliography{Ref_Neuron1}
	
\end{document}